 \documentclass[final,5p,times,twocolumn,authoryear]{elsarticle}

\usepackage{amssymb}
\usepackage{lipsum}
\usepackage{array}
\usepackage{graphicx}
\usepackage{geometry}
\usepackage{amsmath}
\usepackage{graphicx}
\usepackage{natbib}
\setcitestyle{numbers}
\setcitestyle{square}
\newgeometry{left=.8in,right=0.8in,top=1in,bottom=1.3in}

\journal{Electronic Imaging Conference 2025}

\begin{document}

\begin{frontmatter}



\title{Undermining Image and Text Classification Algorithms Using Adversarial Attacks\tnoteref{label1}}
\tnotetext[label1]{The study was conducted during the Next Generation Stem Internship Program 2023 at Oak Ridge National Laboratory}


\author[a,b]{Langalibalele Lunga }
\affiliation[a]{Farragut High School}
\author[b]{Suhas Sreehari}
\affiliation[b]{Oak Ridge National Laboratory}

\begin{abstract}
Machine learning models are prone to adversarial attacks, where inputs can be manipulated in order to cause misclassifications. While previous research has focused on techniques like Generative Adversarial Networks (GANs), there's limited exploration of GANs and Synthetic Minority Oversampling Technique (SMOTE) in text and image classification models to perform adversarial attacks. Our study addresses this gap by training various machine learning models and using GANs and SMOTE to generate additional data points aimed at attacking text classification models. Furthermore, we extend our investigation to face recognition models, training a Convolutional Neural Network(CNN) and subjecting it to adversarial attacks with fast gradient sign perturbations on key features identified by GradCAM, a technique used to highlight key image characteristics of CNNs  use in classification.
Our experiments reveal a significant vulnerability in classification models. Specifically, we observe a 20\% decrease in accuracy for the top-performing text classification models post-attack, along with a 30\% decrease 
in facial recognition accuracy. This highlights the susceptibility of these models to manipulation of input data. Adversarial attacks not only compromise the security but also undermine the reliability of machine learning systems. By showcasing the impact of adversarial attacks on both text classification and face recognition models, our study underscores the urgent need for develop robust defenses against such vulnerabilities. 

\end{abstract}




\end{frontmatter}




\section{Introduction}
\label{introduction}

Machine learning algorithms have experienced an exponential surge in popularity  due to their 
efficiency in making classifications and predictions.
The algorithms have been incorporated into systems that are supporting real world applications, such as object recognition in self driving cars and cancer prediction in medical diagnoses. However, adversarial attacks can make these algorithms insecure and prone to incorrect predictions. An adversarial attack is an input provided to machine learning classifiers for the purpose of causing a misclassification. Past research shows the implications of adversarial attacks in image and text classifiers, demonstrating how adding specific perturbations to inputs result in a substantial decrease in model performance.

In this study, we seek to analyze the types of inputs that fool classification models by utilizing Fast Gradient Sign Method (FGSM) perturbation vectors on the result of GradCAM highlighted features,  GANs, and SMOTE to generate adversarial attacks. As a result, this study demonstrates the vulnerabilities of machine learning models to adversarial attacks using GANs and SMOTE.

This paper presents a novel adversarial attack strategy that combines GANs and SMOTE to target text classifiers and a novel attack on image classifiers with FGSM and GradCAM. Our experiments work to validate the influence of these adversarial attacks against machine learning models deployed in real-world scenarios. The structure of this manuscript is as follows: Section II provides a  review of the existing literature and contributions in the domains of GANs and adversarial attacks. Section III articulates the methodological framework employed in the current investigation. Section IV presents the experimental setup, alongside the resulting data and analysis. Finally, Section V offers a summary of the findings, encapsulates the study's contributions, and outlines potential future research inquiries.

\section{Related Works}
Previous studies center around the use of GANs in computer vision, specifically with image generation and video manipulation. To our knowledge, few studies focus on the use of GANs and SMOTE in tandem for the purpose of adversarial attacks on binary classifiers, along with the combination of GradCAM and FGSM on facial recognition models.

\subsection{Generative Adversarial Networks}
In this paper, we use GANs \citep{goodfellow2020generative}, modeled by  MiniMax loss shown below, to create the adversarial examples used against the financial fraud classifiers.
   
\begin{equation}
    L = \min_{G}\max_{D}\mathbb{E}_{x\sim p_{\text{data}}(x)}[\log{D(x)}] +  \mathbb{E}_{z\sim p_{\text{z}}(z)}[1 - \log{D(G(z))}]
\end{equation}

where $G$ is the generator model, $D$ is the discriminator model, $\mathbb {E}_{x\sim p_{\text{data}}(x)}$ is the distribution of the original dataset, $\log{D(x)}$ is the output of $D$ being maximized,  $\mathbb{E}_{z\sim p_{\text{z}}(z)}$ is the distribution of the noise produced by $G$, and $\log{D(G(z))}]$ is being minimized by $G$.

Generative adversarial networks have many applications, from creating adversarial attacks to generating visually realistic images \citep{sharif2016accessorize} , \cite{kurakin2016adversarial} , \cite{salimans2016improved} .The authors in \citep{sharif2016accessorize} studied the effects that adversarial attacks have on facial biometric systems,investigating novel attacks that allow an attacker to evade recognition or impersonate another individual. They reported that generating accessories in the form of eyeglass frames can effectively fool state-of-the-art face recognition systems. Furthermore, the authors in \cite{kurakin2016adversarial} show the vulnerabilities of classification models  by feeding adversarial images obtained from a cell-phone camera to an ImageNet Inception classifier and measuring the classification accuracy of the system. Their results show that a large fraction of adversarial examples are classified incorrectly even when perceived through the camera. Our research builds upon the methods of utilizing GANs for adversarial attacks by combining it with SMOTE. 
\subsection{Synthetic Minority Oversampling Technique}
Synthetic Minority Oversampling Technique (SMOTE) is an algorithm commonly used to solve class imbalance problems in machine learning fields. Figure 1 shows an example of class imbalance resolved using SMOTE. 

\begin{figure}[h]
    \includegraphics[width=.5\textwidth]{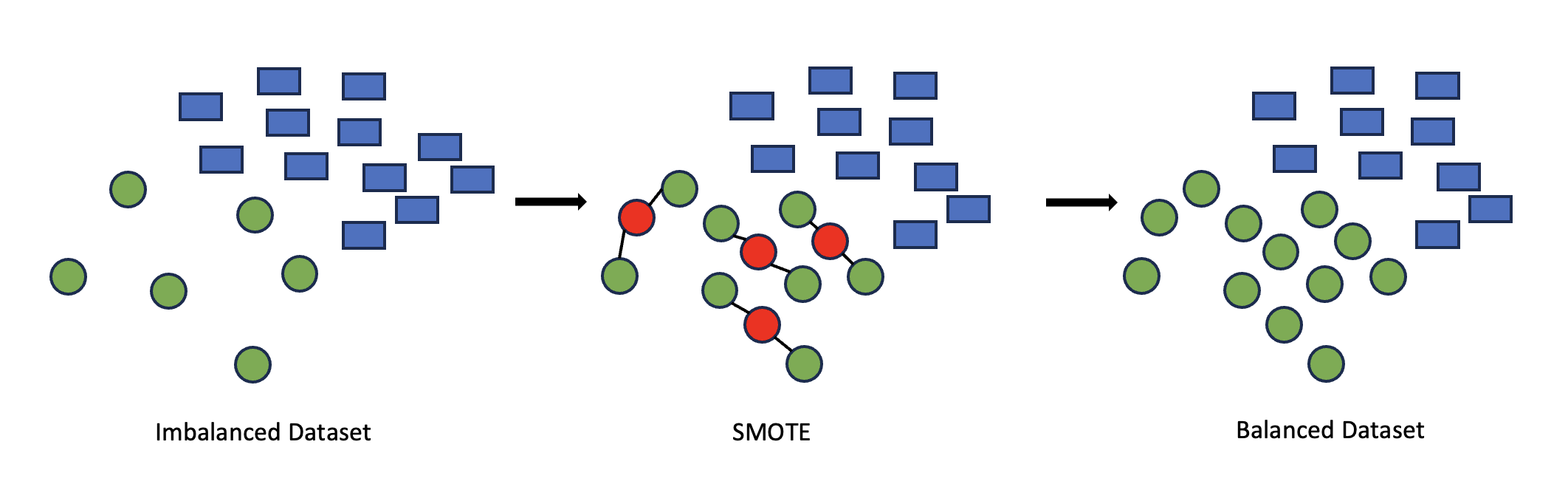}
    \caption{Example SMOTE workflow on class balancing}
\end{figure}

We use SMOTE on the adversarial examples produced by GANs to generate more adversarial examples, or data points that fool the financial fraud classifiers. Specifically, the adversarial examples created by the GANs are treated as a minority class, in which SMOTE is used to create more examples. 

\subsection{Adversarial Attacks/Data Generation with GANs}
Past research has centered the use of GANs in creating deceptive images fed to image classification models. By adding certain perturbations to images,  machine learning models misclassify images with high confidence \cite{goodfellow2014explaining}, \cite{kurakin2016adversarial}, \cite{moosavi2017universal}. For instance, the authors in \cite{moosavi2017universal} propose a systematic algorithm for computing universal perturbations,
and show that state-of-the-art deep neural networks are
highly vulnerable to perturbations imperceptible to the human eye. Researchers have found that GANs can be used to generate tabular/textual data \cite{xu2018synthesizing}, \cite{guo2018long}. The authors in \cite{xu2020adversarial} present Tabular GAN (TGAN), a generative adversarial network which can generate tabular data like medical or educational records. Using TGAN, they generate high-quality and fully synthetic tables while simultaneously generating discrete and continuous variables. The authors in \cite{guo2018long} propose the LeakGAN framework, addressing the low accuracy rates of attempts at generating text of more than 20 words.Their extensive experiments on synthetic data and various real-world tasks demonstrate that LeakGAN is highly effective in long text generation and also improves the performance in short text generation scenarios.

In this study, we utilize the combination of GANs and SMOTE to create synthetic financial data that causes fraud misclassification. We train three machine learning models on a dataset labeled on fraud and non fraud data, and feed it adversarial examples in order to “attack” the models, then recording the impacted accuracy's of each model to measure the effect of the adversarial attack.

\subsection{Carlini Wagner Adversarial Attack}
Another form of adversarial attack is the Carlini Wagner attack developed in \cite{carlini2017towards} , which formulates the generation of adversarial examples as an optimization problem. The attack aims to find a small perturbation \(\delta\) that, when added to input image \(x\), causes the model to misclassify the perturbed image \(x' = x + \delta\). The optimization problem can be described as:

\begin{equation}
\min_\delta \|\delta\|_p + c \cdot f(x + \delta)
\end{equation}

where:
 \(\|\delta\|_p\) is the \(p\)-norm of the perturbation, typically the \(L_2\) norm (\(p=2\)) or \(L_\infty\) norm (\(p=\infty\)), which measures the size of the perturbation.
 \(f(x + \delta)\) is an objective function that represents the degree to which the perturbed image \(x + \delta\) is misclassified by the model.
 \(c\) is a constant that balances the trade-off between minimizing the perturbation's size and maximizing the objective function's value.

The function \(f(x + \delta)\) is designed to encourage the misclassification of the perturbed input. For a targeted attack, it is defined as:

\begin{equation}
f(x + \delta) = \max(\max\{Z(x + \delta)_i : i \neq t\} - Z(x + \delta)_t, -\kappa)
\end{equation}

where \(Z(x + \delta)_i\) denotes the logits (pre-softmax output) of the model for class \(i\)
, \(t\) is the target class that the attacker wants the model to classify the input as.
, \(\kappa\) is a parameter that controls the confidence of the attack; higher values of \(\kappa\) make the attack more confident in the misclassification.
The goal of the optimization is to find the smallest perturbation \(\delta\) that causes the model to classify the input \(x + \delta\) as the target class \(t\) with high confidence. The term \(\|\delta\|_p\) ensures that the perturbation is as small as possible, making the adversarial example harder to detect. The constant \(c\) adjusts the importance of achieving the misclassification relative to keeping the perturbation minimal. 

This formulation makes the Carlini Wagner attack particularly powerful because it produces perturbations that are often imperceptible to humans but effective in misleading the model. While the Carlini Wagner attack is highly effective, we use the Fast Gradient Sign attack due to its low computational cost and efficiency.

\subsection{Gradient-Weighted Class Activation Mapping}
Gradient Class Activation Mapping refers to a technique used in order to highlight the features that object and image recognition algorithms focus on in order to make a classification. GradCAM is frequently used in tandem with CNNs to enhance model transparency and provide insights into how and what the model relies on to make certain predictions. Specifically,  GradCAM utilizes the gradients of the last layers in a CNN in order to create a localization map highlighting the important regions in the image for predicting the concept \cite{selvaraju2017grad}. Figure 2 demonstrates the result of applying GradCAM on the Olivetti Faces dataset:
\begin{figure}[h]
    \centering
    {\includegraphics[width=0.1\textwidth]{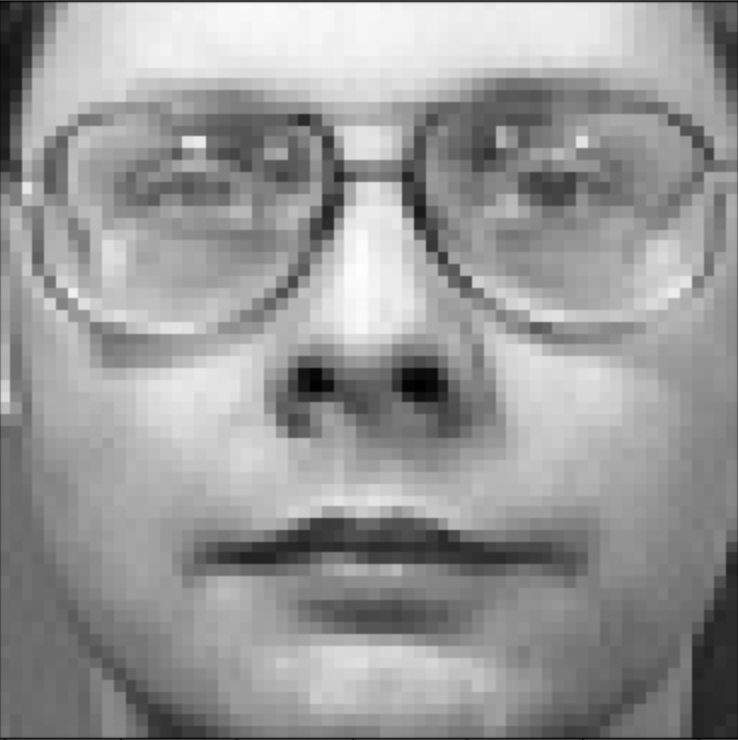}} 
    {\includegraphics[width=0.1\textwidth]{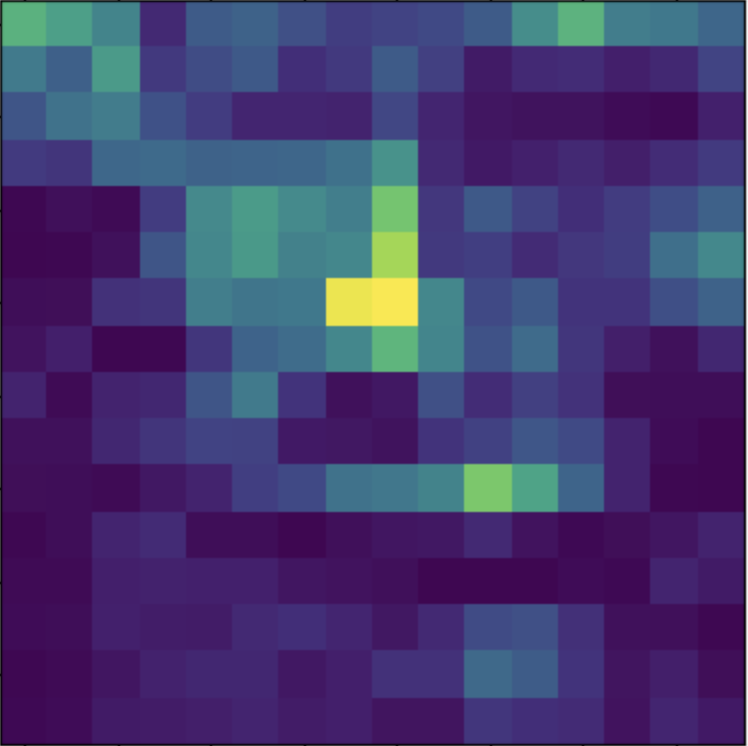}} 
    {\includegraphics[width=0.1\textwidth]{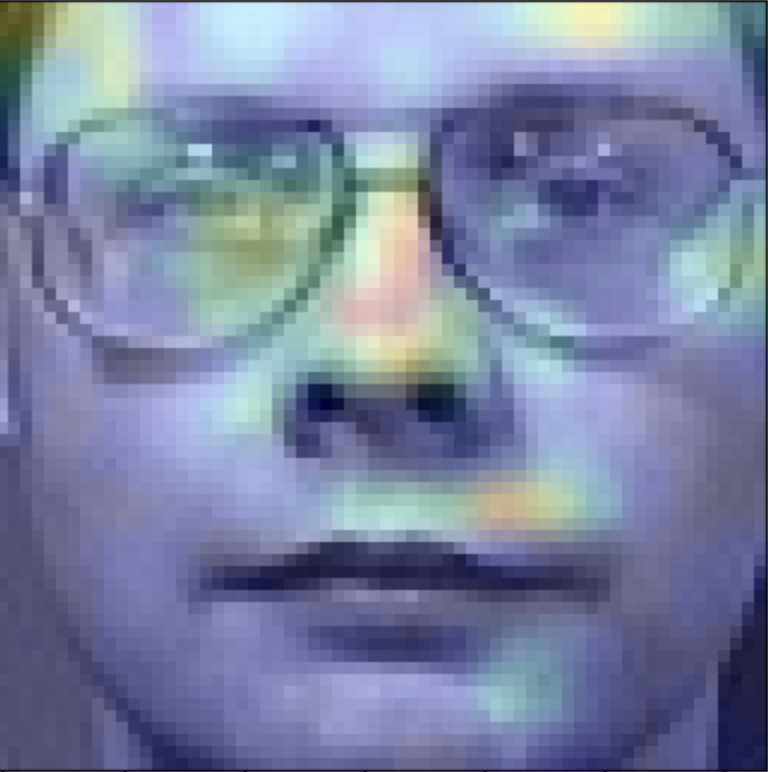}}
\end{figure}

\begin{figure}[h]
    \centering
    {\includegraphics[width=0.1\textwidth]{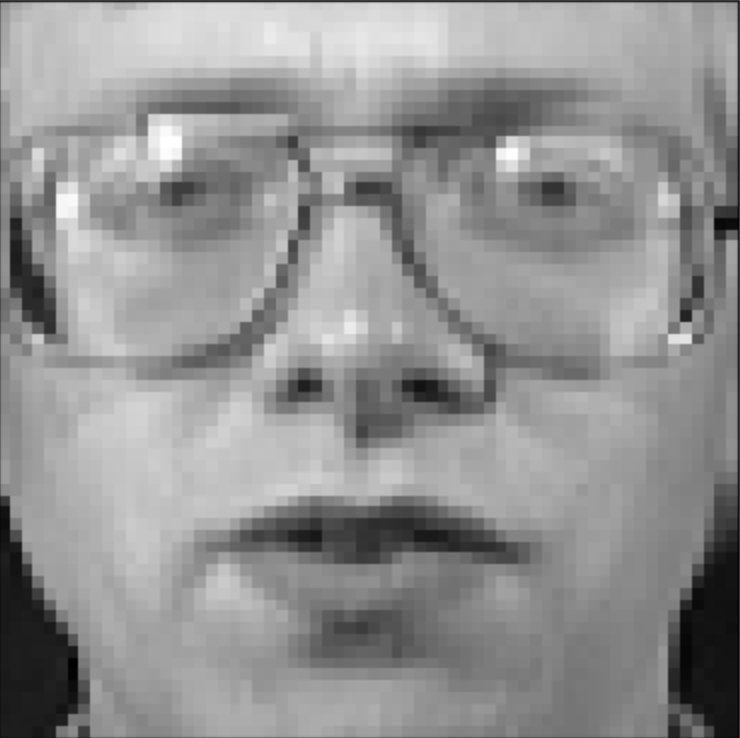}} 
    {\includegraphics[width=0.1\textwidth]{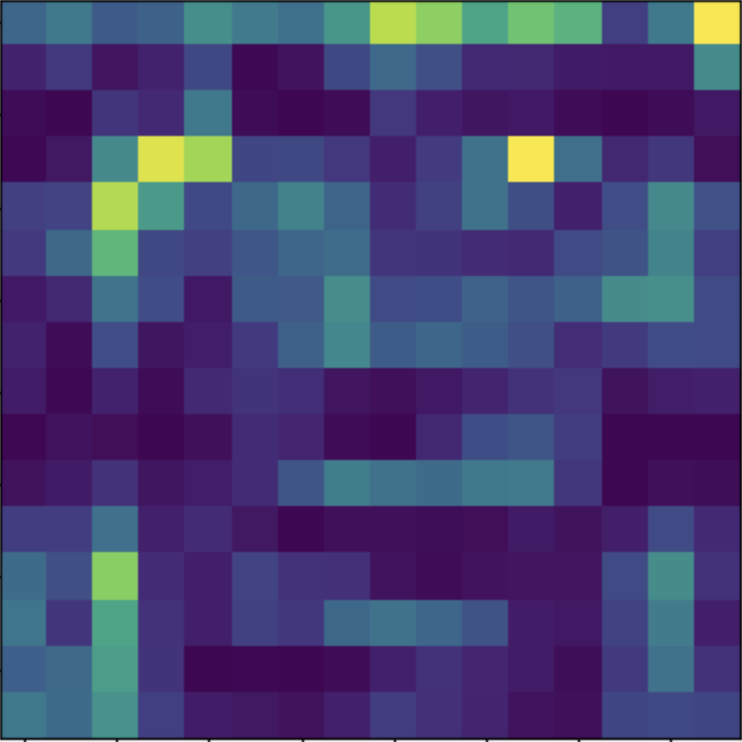}} 
    {\includegraphics[width=0.1\textwidth]{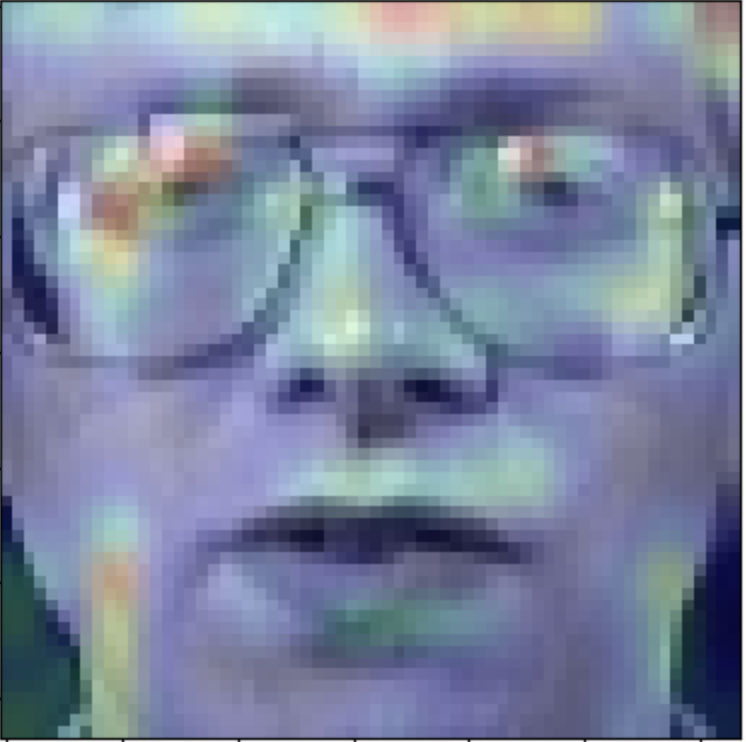}}
    \caption{First column are two input examples from the Olivetti Faces dataset (Row 1: Person 3 and Row 2 is Person 2). Second column are GradCAM heatmap of the activation maps. Third column shows the corresponding input GradCAM features that are key for CNN predictions.}
    
\end{figure}

\subsection{Fast Gradient Sign Method}
Fast Gradient sign method(FGSM)\cite{goodfellow2014explaining} is an adversarial attack that perturbs input data into a model using noise that is based off of the gradient of the loss with respect to the input. FGSM can be modeled by the equation below:
\begin{equation}
    \eta = \epsilon * sign (\nabla_ x \textsl{J}(\theta,x,y))
\end{equation}

where $\eta$ is the perturbation added to an image, $\epsilon$ is the strength of perturbation, sign $\nabla_x$ is the sign of the gradient with respect to the image, $x$ is the input and $y$ is the output
FGSM adds a visually indistinguishable perturbation that fools the model using a certain value of epsilon. The gradient is calculated with respect to the inputs of the model in order to maximize the loss. Higher values of epsilon result in more visible and larger perturbations while lower values are more subtle. FGSM is commonly used to showcase the vulnerability of deep learning models and investigate the transferability of adversarial attacks, however, this research study takes a different approach by combining it with GradCAM. Specifically, GradCAM is used to highlight important features, and FGSM is applied to noted features to implement the adversarial attack.

\section{Methodology}
This section outlines the comprehensive methodology employed in this paper, with a focus on demonstrating the susceptibility of image and text classification models to adversarial attacks.

The rationale for our approach is as follows:
Generative Adversarial Networks (GANs) are used to generate adversarial examples due to their capacity to mimic the underlying distribution of input data, leading to highly effective adversarial attacks. GANs utilize a generator and discriminator in a minimax game where the generator produces fake data, and the discriminator attempts to distinguish it from real data. This iterative process helps generate imperceptible adversarial perturbations. SMOTE, on the other hand, is used to address class imbalance, which can be a significant issue in adversarial data generation. By applying SMOTE to the misclassified instances (treated as the minority class), we are able to over sample and generate more adversarial samples, further enhancing the adversarial attack’s impact.

For the text classification models, the  data points that the generated data from the GANs and SMOTE is based off on are the data points that all three machine learning models misclassified. We create data points similar to these because these data points are ones that rely on the decision boundary of our three trained fraud detectors, serving as prime examples of scenarios that our fraud classifiers would have trouble identifying. 

\subsection{Boundary point generation with SMOTE}

SMOTE fixes imbalance datasets through statistical means, specifically through the feature space of each target class and its nearest neighbors. SMOTE is preceded by two commonly known methods of handling imbalanced datasets: under sampling and oversampling. 

Over sampling and under sampling are prevalent techniques used to address class imbalance in datasets. Oversampling involves increasing the number of instances in the minority class by duplicating or generating synthetic data points. This method aims to balance the representation of classes by augmenting the smaller class to match the larger one. Conversely, under sampling reduces the number of instances in the majority class to create a more balanced dataset by randomly eliminating samples from the overrepresented class. SMOTE stands out by creating synthetic instances that are strategically generated based on the feature space of the minority class and the proximity of its nearest neighbors. Unlike traditional oversampling methods that merely replicate existing instances, SMOTE generates new data points that maintain the underlying characteristics of the minority class, thereby enhancing the overall balance of the dataset without losing crucial information. 
This is done using the $l2$ norm modeled by the equation:
\begin{equation}
    d = ||x - y||_2 ,\, where\,x  = (x_{i})_{i=0}^{n-1} \, , y= (y_{i})_{i=0}^{n-1}
\end{equation}

For the purposes of this study, SMOTE is used to create new instances based off of the data the three models incorrectly classified. The dataset of test cases all models failed to classify is treated as a dataset with imbalance. Specifically, this dataset has 139 instances of fraud and 14 instances of non-fraud. Using SMOTE, this dataset is resampled to balance the classes out, with 139 instances of fraud and non-fraud. This re-sampled data results in a dataset with adversarial instances that have been generated for later use in tandem with the results from a GAN.
\subsection{Boundary point generation with GANs}
GANs learn the probability distribution of a dataset, and then use the estimated distribution to generate more examples\cite{goodfellow2020generative}. Figure 3 shows the data generation process using GANs. 
\begin{figure}[h]
          \makebox[\textwidth][l]{\includegraphics[width=.4\textwidth]{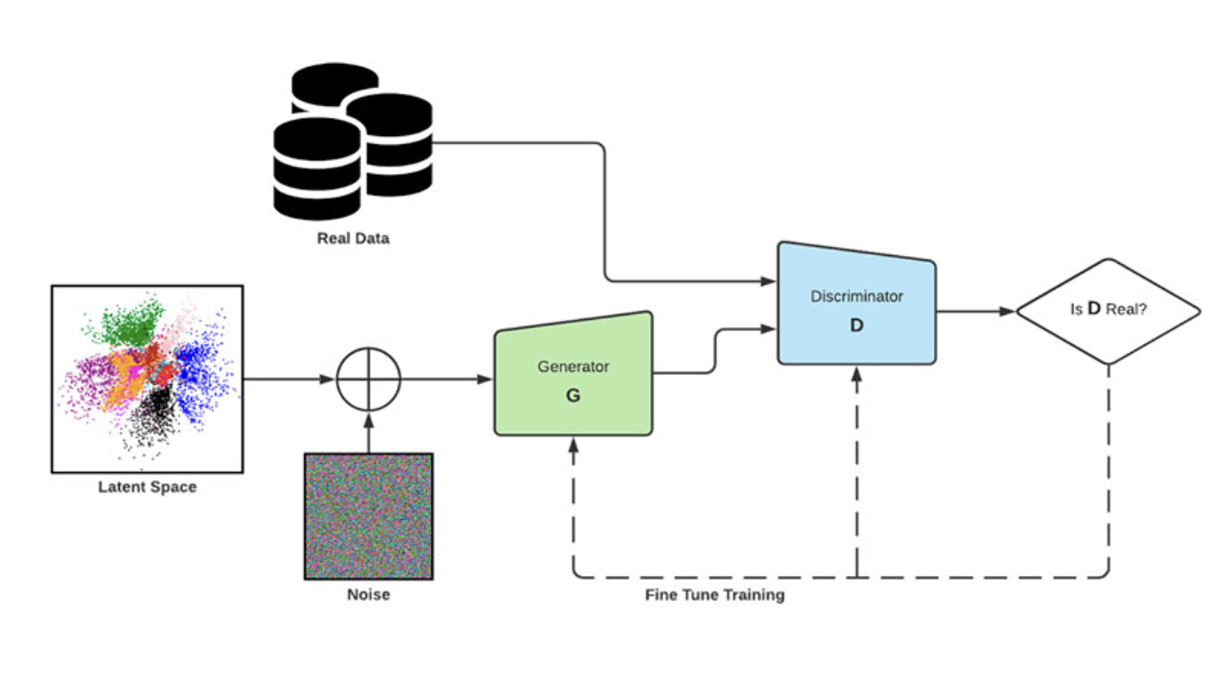}}
    \caption{Workflow for tabular data generation using GANs.}
    \label{fig:enter-label}
\end{figure}

In this study, we use a specific variant of  GAN with the intent of analyzing and generating tabular data points: Conditional Tabular Generative Adversarial Network(CTGAN)\cite{xu2019modeling}. CTGAN is designed to accurately model and generate instances of data where continuous columns have numerous modes and categorical columns are imbalanced. CTGAN uses several GAN techniques, namely conditional generators and training by sampling. Compared to GAN, CTGAN allows for the structured generation of data in a tabular format. CTGAN, in addition, preserves the statistical distribution of the tabular data, whereas GAN would fail to recognize instances of class imbalance\cite{xu2019modeling}. We implemented CTGAN on the previous dataset that contained all values that the three machine learning models failed to classify from the test cases. We utilize CTGAN with several default parameters, namely epochs, batch size, generator/discriminator dimensions, and noise embedding dimensions.

\begin{center}

\begin{table}[h]
    
\caption{Parameters of CTGAN model}
\resizebox{\columnwidth}{!}{%
\begin{tabular}{|c||c c c c|} 

 \hline
 Epochs & batch\_size & gen\_dim & discrim\_dim & noise\_embedd  \\
 \hline
 100 & 500 & (256,256) & (256,256) & 128\\
 \hline
\end{tabular}
}

\end{table}
\end{center}

\subsection{GradCAM and Perturbations with FGSM}

Following the training of the CNN, we begin the process of the adversarial attack with a GradCAM, a powerful method of describing what the CNN is using to make classifications. GradCAM is implemented on each face in the test dataset, and the resulting heatmap is superimposed. In order to perturb the most important features on each face, we threshold the result of each heatmap to a value of .4, with values below .4 being 0 and all others being 1. We then generate a targeted adversarial attack to apply on the threshholded  heatmap.  The creation of the adversarial attack is dependent on 4 parameters: model(CNN), baseImage, delta(noise vector), steps, and a class label we want the model to incorrectly predict. The algorithm used to generate the noise vectors and create the adversarial attack utilizes gradient descent in order to minimize the probability that the model predicts the correct class while maximizing the probability that the model selects the targeted class, hence the term targeted attack.

Specifically, we add a perturbation vector to the image being manipulated and preprocess the result. The result is then fed to the CNN and the categorical cross-entropy loss with respect to the both the original class label and the target class label is calculated. This process is repeated 500 times, utilizing gradient descent to minimize the probability that the model predicts the correct class while maximizing the probability that the model selects the targeted class.



\section{Experiments and Results}

This section provides a detailed explanation of the data, experimental results, and discussions. 

\textbf{Financial Fraud Data:} The first set of experiments employ a comprehensive dataset designated for the analysis of bank account fraud, which was sourced from the Kaggle platform \citep{jesus2022turning}. The dataset is expansive, encompassing 1,000,000 instances each described by 32 distinct attributes. Notably, each instance is annotated with a binary label indicating the presence of fraudulent or legitimate account activity. This dataset is characterized by a heterogeneous mix of attribute types, including both textual and numerical data. In order to run the data through our models, the textual data required conversion to numerical data. For this purpose, we integrated the Label Encoding technique. This preprocessing step is critical for the transformation of categorical attributes—specifically, those pertaining to payment type, employment status, housing status, and the operating system of the device—into a numerical representation. In order to decrease run times, a subset 20,000 from the original 1,000,000 rows of data was used before the implementation of SMOTE, CTGAN, and training of classifiers.

\textbf{Training of Text Classification Models and Finding Boundary Points:} Our purpose in training these models is to simulate real world machine learning algorithms that detect financial fraud, our goal being to show their vulnerabilities to adversarial attacks. Three machine learning models are trained: DecisionTree, Random Forest, and XGradientBoosting.  However, due to the imbalanced nature of the dataset, the original dataset is split into all rows with fraud and all rows without fraud. From the original dataset, there are 988971 rows identified as non- fraudulent and 11029 rows identified as fraudulent. Training the models on an imbalanced dataset would result in bias towards the majority class, in this case non-fraudulent instances. To counter this, a subset of the non-fraudulent data instances was created containing 10,000 rows, equal to the number of fraudulent instances from the original data set. Next, this subset of non-fraudulent instances was combined with the fraudulent instances in the original dataset to create a dataset with an equal amount of each class. Due to the mix of categorical and numerical values, LabelEncoding was performed on our new, balanced dataset. This dataset is then prepped for model training, where we utilize sklearn to create an 8:2 ratio between the training and test data. The three models are all trained on this equalized dataset and metrics are recorded. 

Following the training and testing stage, the incorrect predictions for all three models are collected by comparing each instance of a prediction to its true value from testing. An empty dataframe is created, and data instances that all three models fail to correctly classify are appended to said dataframe. This dataset of incorrect values serves as the boundary points across all three models of data points that cause confusion in classification. Throughout the study, we use this dataset as the input to SMOTE and GAN to generate synthetic data that is similar to boundary points.

\textbf{Results on Financial Fraud Detection:}
The initial metrics from training the three financial fraud classification models on a balanced dataset is as follows:
\begin{center}
\begin{table}[h]
\caption{ Performance metrics before adversarial attack}
\resizebox{\columnwidth}{!}{%
\begin{tabular}{|c||c |c| c| c||} 
 \hline
  Metrics &  Decision Tree & Random Forest & XGB \\
 \hline
 Accuracy & 91.22\% & 94.175\% & 94.12\%  \\
 AUC & 0.91 & 0.94 & 0.94  \\
 Recall & 0.91 & 0.89 & 0.91  \\
 Precision & 0.91 & 0.99 & 0.97\\
 
 \hline
\end{tabular}
}
\label{Table1}
\end{table}
\end{center}
In addition, a table showing the true and false positive/negatives for each model is shown below based off their performance on the dataset. Like the results above, this table displays the results from the initial train/test on the balanced dataset.

\begin{center}
\begin{table}[h]
\caption{Classification Results before adversarial attack}
\resizebox{\columnwidth}{!}{%
\begin{tabular}{|c||c |c| c| c||} 
 \hline
  Metrics &  Decision Tree & Random Forest & XGB \\
 \hline
 True Positives & 1822 & 1977 & 1946  \\
True Negatives & 1827 & 1801 & 1837  \\
False Positives & 165 & 191 & 155  \\
False Negatives & 186 & 31 & 62 \\ 
 \hline
\end{tabular}
}
\end{table}
\end{center}

The table below shows the effects of the adversarial attack on the fraud classifiers:

\begin{center}
\begin{table}[h]
\caption{Performance metrics after adversarial attack}

\resizebox{\columnwidth}{!}{%
\begin{tabular}[ht]{|c||c |c| c| c||} 
 \hline
  Metrics &  Decision Tree & Random Forest & XGB \\
 \hline
 Accuracy & 68.75\% & 62.11\% & 65.94\%  \\
 AUC & 0.73 & 0.69 & 0.72  \\
 Recall & 0.58 & 0.45 & 0.53\\ 
 Precision & 0.89 & 0.93 & 0.92\\
 \hline
\end{tabular}
}
\end{table}
\end{center}

Similar to the results above, this table shows the true and false positive/negatives for each model after applying the adversarial attacks. 

\begin{center}
\begin{table}[h]
\caption{Classification Results after adversarial attack}
\resizebox{\columnwidth}{!}{%
\begin{tabular}{|c||c |c| c| c||} 
 \hline
  Metrics &  Decision Tree & Random Forest & XGB \\
 \hline
 True Positives & 1211 & 1304 & 1255  \\
True Negatives & 1535 & 1177 & 1379  \\
False Positives & 1074 & 1432 & 1230  \\
False Negatives & 174 & 81 & 130 \\
 \hline
\end{tabular}
}
\end{table}
\end{center}

\textbf{Facial Recognition Dataset:} The Olivetti Faces Dataset, sourced from AT \& T Laboratories Cambridge, consists of 400 images of 40 different men and women.

\begin{figure}[h]
    \centering
    {\includegraphics[width=0.3\textwidth]{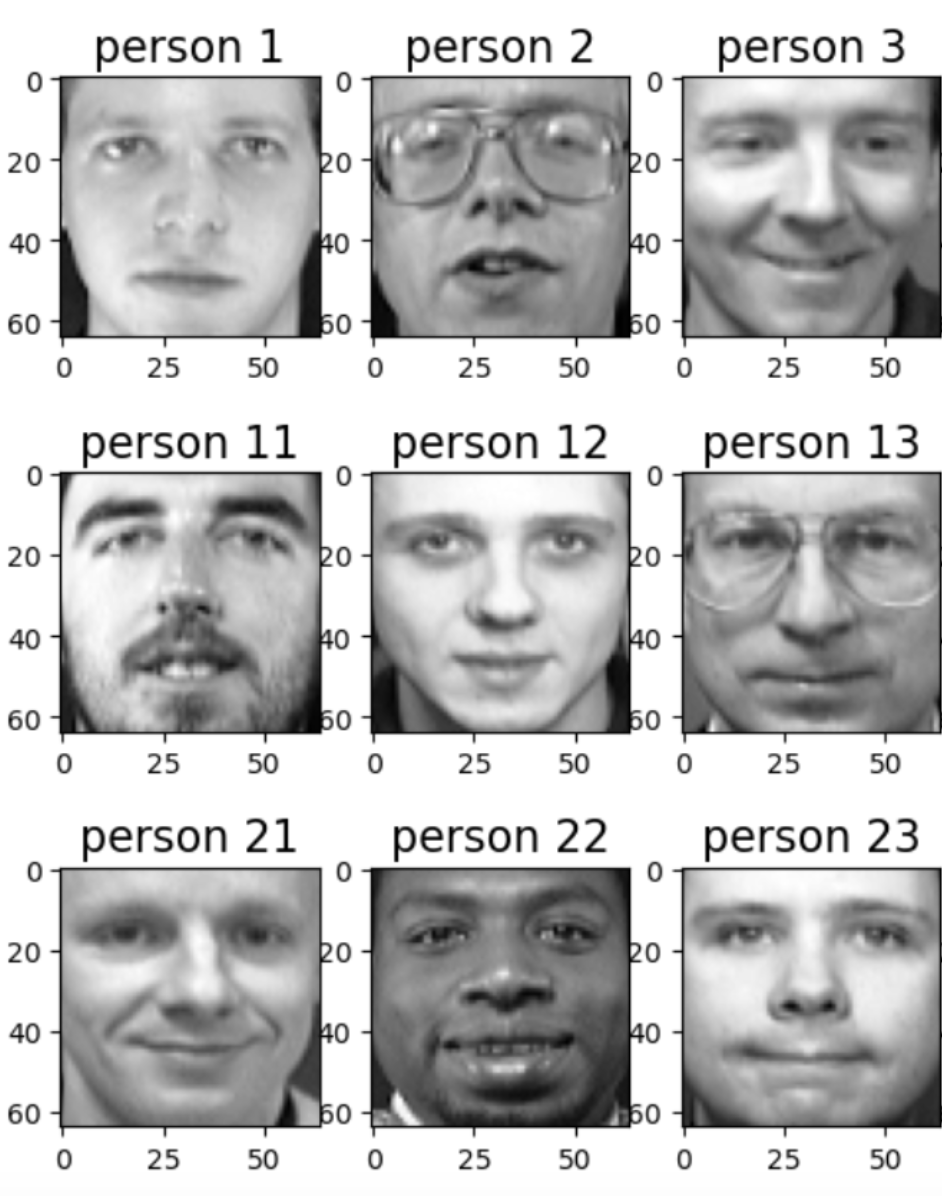}} 
    
    \caption{ Input example faces and their true labels from the Olivetti dataset}
    \label{fig:foobar}
\end{figure}
Each image in the dataset has dimensions of $64 \times 64$ pixels. The dataset is structured so that every 10 instances represent the same person, but in a different lighting, angle, or facial expression.

\textbf{Facial Recognition Training:} Our goal in training a CNN for facial recognition is similar to our motivation with the financial fraud classifiers- we seek to simulate how a real world biometrics system can be fooled by adversarial attacks, creating security risks.
For training the CNN , we use a 20:80 test train split, taking the first 2 images of every 10 in the dataset and using them for test, and the rest for training. Model parameters were imperatively chosen for their  efficiency and high accuracy.
The input of the CNN is an array of dimensions $1 \times 64 \times 64 \times 1$ and the model architecture and training is as follows: \\
\begin{figure}[h]
    \centering
    {\includegraphics[width=0.45\textwidth]{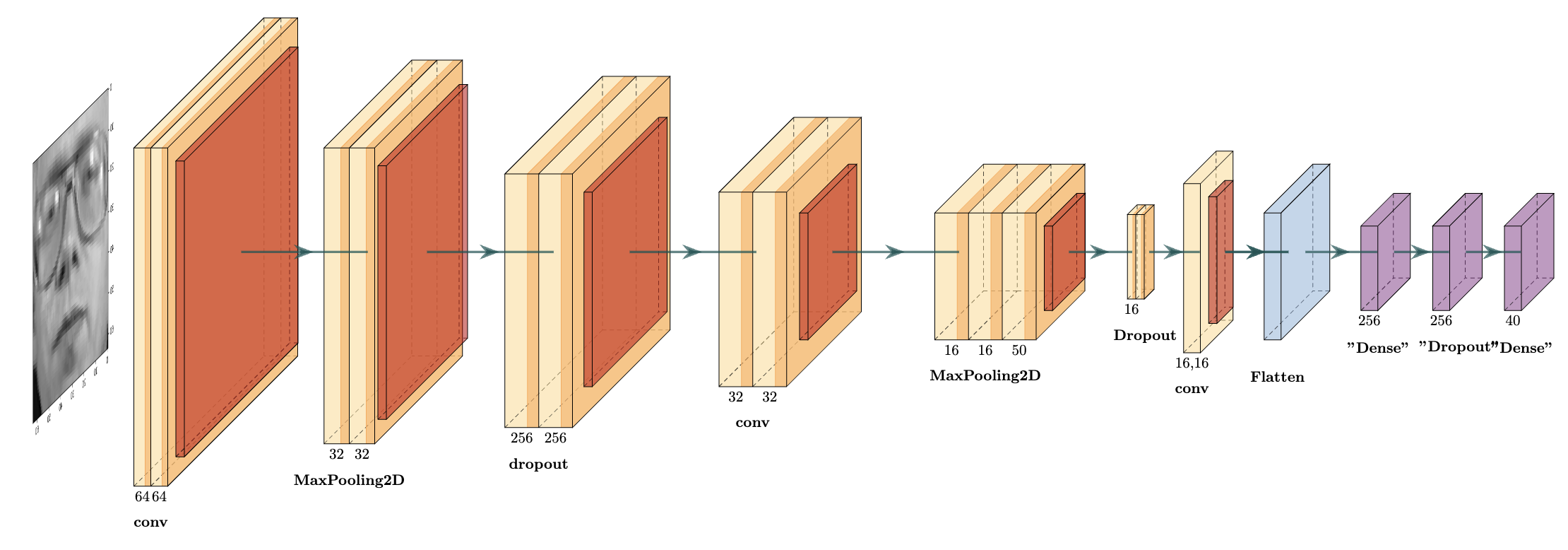}} 
    \caption{CNN model architecture.}
    \label{fig:foobar}
\end{figure}

\textbf{Results on Facial Biometric Systems}
The initial accuracy of the CNN was 98.75\%. However, the adversarial attack dropped accuracy to 68\%.  Below is a visual of the three results. 

\begin{table}[h]
\centering
\caption{CNN experimental results with adversarial attacks: Column 1 shows the true label; Columns 2 and 3 display the original and perturbed Olivetti faces; Columns 4 and 5 show model predictions before and after the attack; Column 6 displays the misclassified face matching post-attack prediction.}

\vspace{0.5em}

\resizebox{.48\textwidth}{!}{
\begin{tabular}{|c|c|c|c|c|c|}
\hline
\textbf{True Label} & \textbf{Olivetti Face} & \textbf{Perturbed Inputs} & \textbf{Prediction Before Attack} & \textbf{Prediction with Attack} & \textbf{Incorrect Face Example} \\
\hline
16 & \includegraphics[width=2.5cm, height=2.2cm]{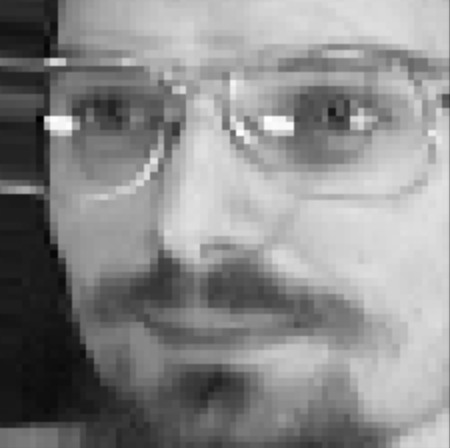} & \includegraphics[width=2.5cm, height=2.2cm]{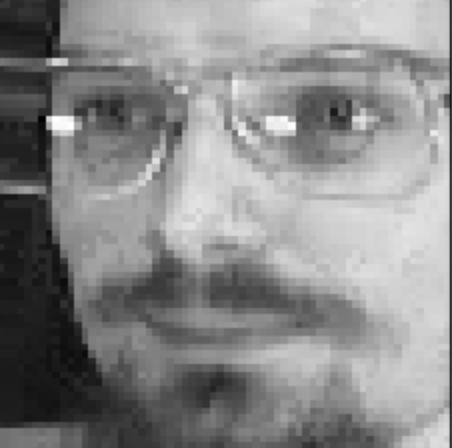} & Person 16 & Person 2 & \includegraphics[width=2.5cm, height=2.2cm]{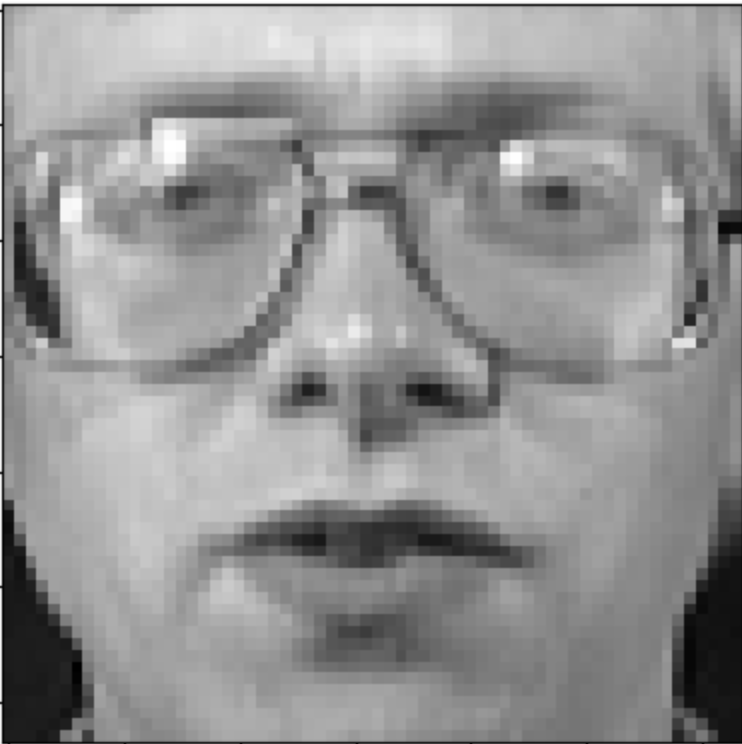} \\
\hline
22 & \includegraphics[width=2.5cm, height=2.2cm]{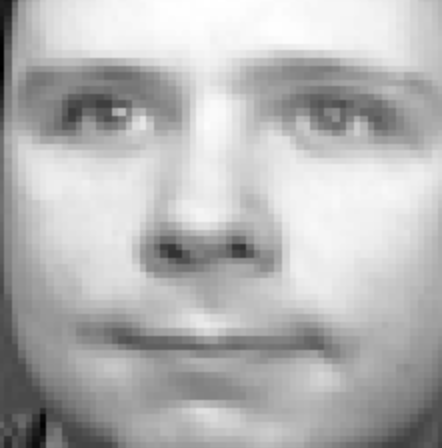} & \includegraphics[width=2.5cm, height=2.2cm]{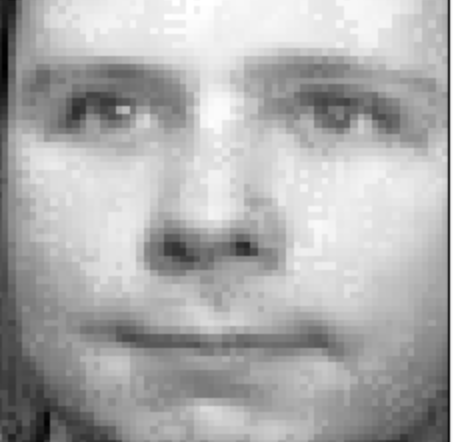} & Person 22 & Person 26 & \includegraphics[width=2.5cm, height=2.2cm]{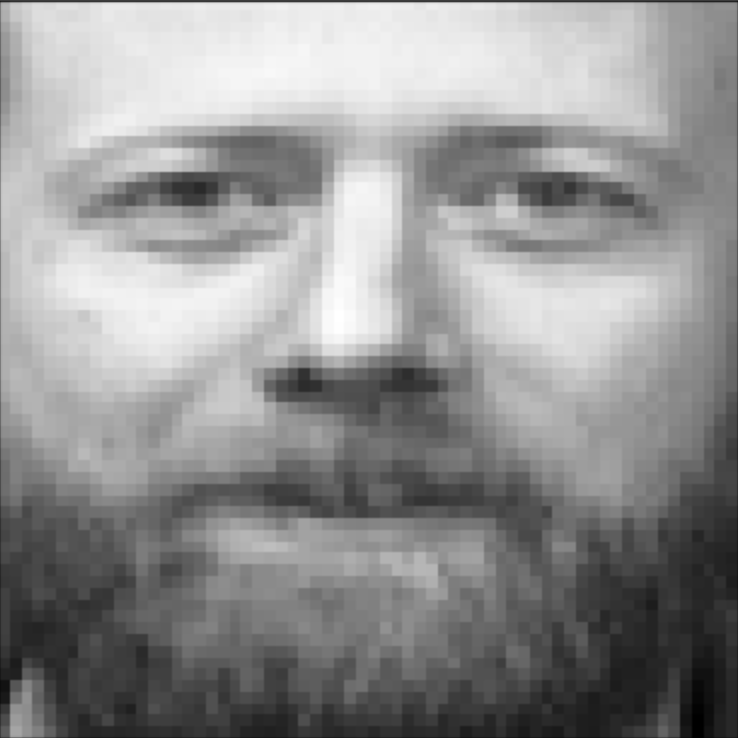} \\
\hline
35 & \includegraphics[width=2.5cm, height=2.2cm]{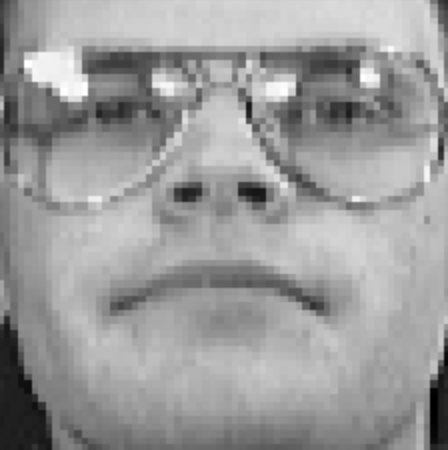} & \includegraphics[width=2.5cm, height=2.2cm]{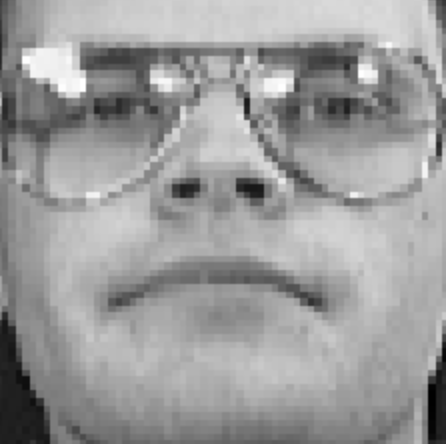} & Person 35 & Person 7 & \includegraphics[width=2.5cm, height=2.2cm]{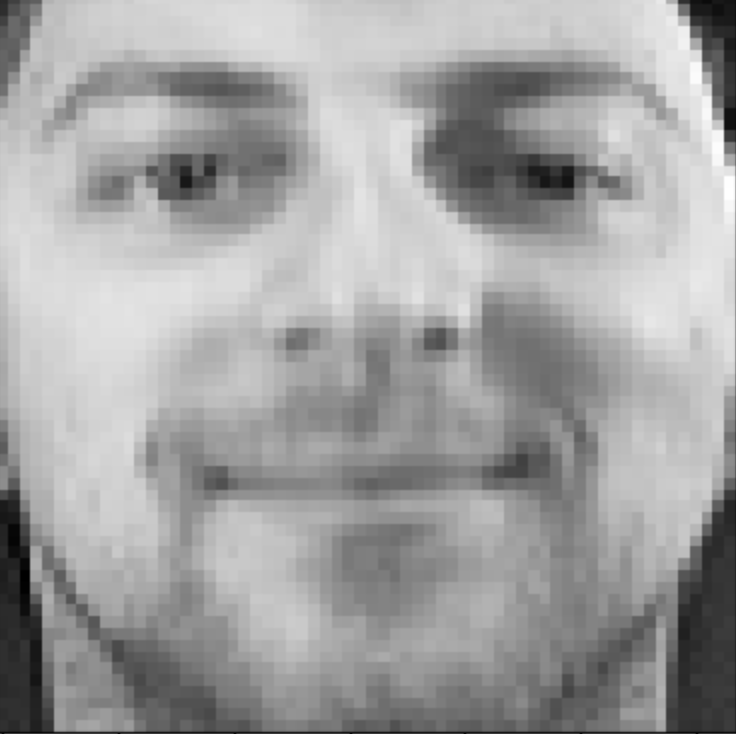} \\
\hline
\end{tabular}
}
\end{table}

\section{Conclusions}
The results demonstrate a significant decrease in model performance following adversarial attacks. In particular, the text classification models experienced a 20\% drop in accuracy, while facial recognition accuracy dropped
by 30\% . This suggests that image classifiers, especially those relying on CNNs, may be more vulnerable to adversarial perturbations compared to text classifiers. One possible explanation for this difference is that CNNs rely heavily on local pixel-based patterns, which can be easily disrupted by small, imperceptible changes to the input data, as highlighted by the FGSM attack. 

In contrast, text classification models often rely on semantic structures, which may require more sophisticated perturbations to achieve a similar impact. This suggests that adversarial defenses may need to be more rigorously developed for image-based models. Furthermore, the drop in accuracy might indicate that even well-performing models with high initial accuracies are not immune to adversarial attacks, calling into question the reliability of these models in high-stakes real-world applications such as financial fraud detection and biometric security systems. 

The significant drop in performance after introducing adversarial examples highlights an urgent need for robust adversarial defense mechanisms, such as adversarial training or input sanitization techniques, to mitigate these vulnerabilities. Future research should explore adversarial training and robust defensive strategies, such as adversarial noise detection, to mitigate these vulnerabilities.



\nocite{*}
\bibliographystyle{acm} 

\bibliography{references}


\end{document}